\documentclass[prl,superscriptaddress,twocolumn]{revtex4-1}
\usepackage{amsmath,amssymb,mathrsfs}
\usepackage{graphicx}
\usepackage{rotating}
\usepackage{amssymb}
\usepackage[usenames,dvipsnames]{color}
\usepackage[
colorlinks=true,
linkcolor=blue,
citecolor=BrickRed,
urlcolor=blue]{hyperref}
\usepackage[all]{hypcap}
\graphicspath{ {./figs/} }
\usepackage{listings}
\usepackage{tikz}

\tikzset{middlearrow/.style={
    decoration={markings,
      mark= at position 0.55 with {\arrow[scale=1,blue]{#1}} ,
    },
    postaction={decorate}
  }
}
\usetikzlibrary{decorations.pathreplacing,
  decorations.markings,decorations.pathmorphing}
\usetikzlibrary{calc}

\begin{document}
\title{Competing many-body instabilities in two-dimensional dipolar Fermi 
  gases}
\author{Ahmet Kele\c{s}}
\affiliation{Department of Physics and Astronomy, 
  University of Pittsburgh, Pittsburgh, PA 15260}
\affiliation{Department of Physics and Astronomy,
  George Mason University, Fairfax, VA 22030}
\author{Erhai Zhao}
\affiliation{Department of Physics and Astronomy,
  George Mason University, Fairfax, VA 22030}

\begin{abstract}  
  Experiments on quantum degenerate Fermi gases of magnetic
  atoms and dipolar molecules begin to probe their broken symmetry phases
  dominated by the long-range, anisotropic dipole-dipole interaction.  Several
  candidate phases including the $p$-wave superfluid, the stripe density wave,
  and a supersolid have been proposed theoretically for two-dimensional
  spinless dipolar Fermi gases. Yet the phase boundaries predicted by
  different approximations vary greatly, and a definitive phase diagram is
  still lacking. Here we present a theory that treats all competing many-body
  instabilities in the particle-particle and particle-hole channel on equal
  footing. We obtain the low temperature phase diagram by numerically solving
  the functional renormalization-group flow equations and find a nontrivial
  density wave phase at small dipolar tilting angles and strong interactions,
  but no evidence of the supersolid phase. We also estimate the critical
  temperatures of the ordered phases.  
\end{abstract} 

\maketitle

Fermi gases and Fermi liquids play a fundamental role in many-body physics.
Many archetypical broken symmetry phases ranging from superconductivity,
charge density waves to quantum liquid crystals may be understood as
instabilities of an underlying Fermi liquid in a particular interaction
channel. Historically, electron gas with Coulomb interaction, liquid
helium-3, and ultracold Fermi gases of alkali atoms with contact interaction
have served as the testing grounds for many-body theories. Recent experiments
have ushered in a new class of interacting Fermi gases --- the quantum
degenerate gases of fermionic atoms with large magnetic moments such as
$^{161}$Dy  \cite{PhysRevLett.108.215301}, $^{167}$Er
\cite{PhysRevLett.112.010404}, and  $^{53}$Cr \cite{PhysRevA.91.011603} and
ground-state polar molecules such as $^{40}$K$^{87}$Rb  \cite{ni,ospel} and
$^{23}$Na$^{40}$K \cite{PhysRevLett.114.205302,njp-zw,PhysRevLett.109.085301}.
Their low temperature phases are dictated by the dipole-dipole interaction
which is long-ranged, anisotropic, and attractive or repulsive
depending on the relative orientation of the two dipoles. This unique
interaction gives rise to a rich variety of interesting quantum phases
\cite{baranov2012condensed,Baranov-rev,yi-wu-review}.

Take the single-species (spinless) dipolar Fermi gas confined in two
dimensions (2D) for example. Previous theoretical work has identified two
broken symmetry phases. A density wave (DW) is shown to develop when
the dipolar interaction is repulsive, e.g. when the dipoles are aligned normal
to the 2D plane by the external field, and sufficiently strong
\cite{PhysRevA.82.013643,PhysRevB.84.235124,PhysRevLett.108.145304,
  PhysRevB.82.075105,PhysRevB.90.155102,zinner-NJP,PhysRevA.92.023614}.  It features a periodic modulation
of particle density in the form of unidirectional stripes. When the dipoles
are tilted toward the plane beyond a critical angle, the dipolar interaction
becomes partially attractive and supports Cooper pairing \cite{you}. A broad
region of Bardeen-Cooper-Schrieffer (BCS) superfluid phase with $p$-wave
symmetry was predicted \cite{PhysRevLett.101.245301,baran-a,PhysRevA.81.063642}. In
the limit of large tilting angle and strong attraction, the system becomes
unstable: the compressibility becomes negative and the gas is believed to collapse
\cite{PhysRevA.82.013643,PhysRevLett.108.145304,PhysRevA.84.063633,
PhysRevLett.101.245301}. 
While the qualitative picture of the competing DW and BCS instabilities is
agreed upon, there is yet a consensus on a definitive phase diagram. For
example, the DW instability is predicted to occur when the dimensionless
interaction strength [defined below Eq. \eqref{eq:dipolarInteraction}] $g_c=1.45$ within the conserving
Hartree-Fock (HF) approximation \cite{PhysRevA.84.063633,PhysRevB.84.235124}.
By contrast, the Random Phase Approximation (RPA) gives $g_c\sim 0.7$
\cite{PhysRevB.82.075105,PhysRevA.82.013643}. Ref.
\cite{PhysRevLett.108.145304} improved RPA by incorporating exchange
correlations to find a considerably larger $g_c\sim 6$. The fixed-node
Monte-Carlo calculation of Ref. \cite{PhysRevLett.109.200401} however did not
find any evidence for the stripe phase. Moreover, mean field theory suggests a
supersolid phase, i.e. the coexistence of the BCS and DW order, in a finite
region of the phase diagram \cite{PhysRevB.91.224504}. It remains unclear
however whether the supersolid phase can survive quantum fluctuations.

These discrepancies and open questions highlight the challenges to develop an
accurate theory for 2D dipolar Fermi gas. Ideally, the theory should have the
following capabilities: (1) It keeps track of all many-body instabilities in
the particle-particle and the particle-hole channel including the subtle
interplay of the BCS and DW order for intermediate dipole tilting angles. (2)
It extracts the momentum-dependent {effective interactions} between
quasiparticles \cite{PhysRevA.85.023614,wu-sarma} on the Fermi surface
systematically from the bare dipolar interaction. (3) It takes into account
thermal fluctuations to yield phase diagrams at low temperatures of interest
to experiments. (4) It describes quantum fluctuations beyond HF and RPA.

\begin{figure}
  \centering 
  \includegraphics[width=0.47\textwidth]{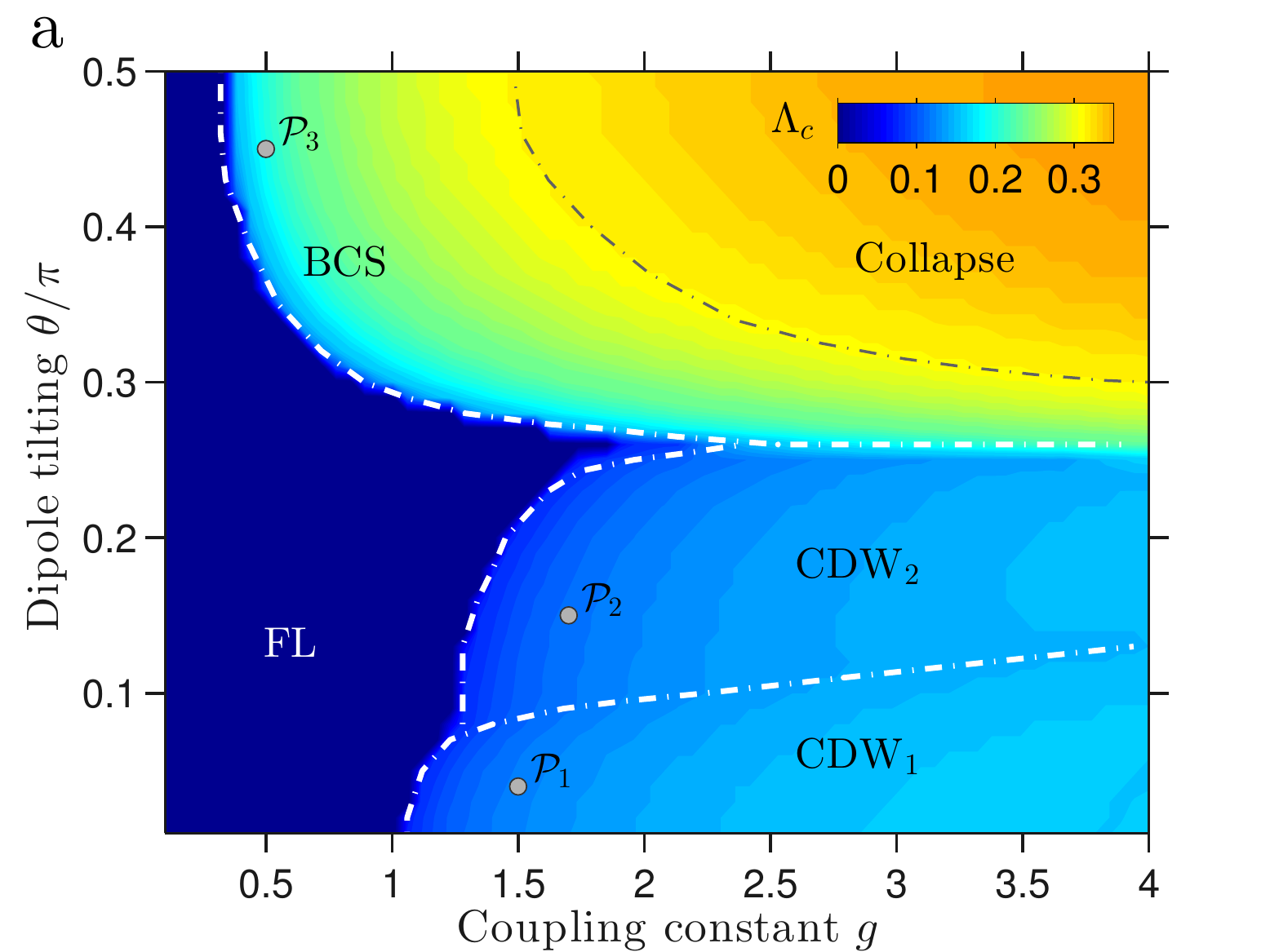}
  \includegraphics[width=0.47\textwidth]{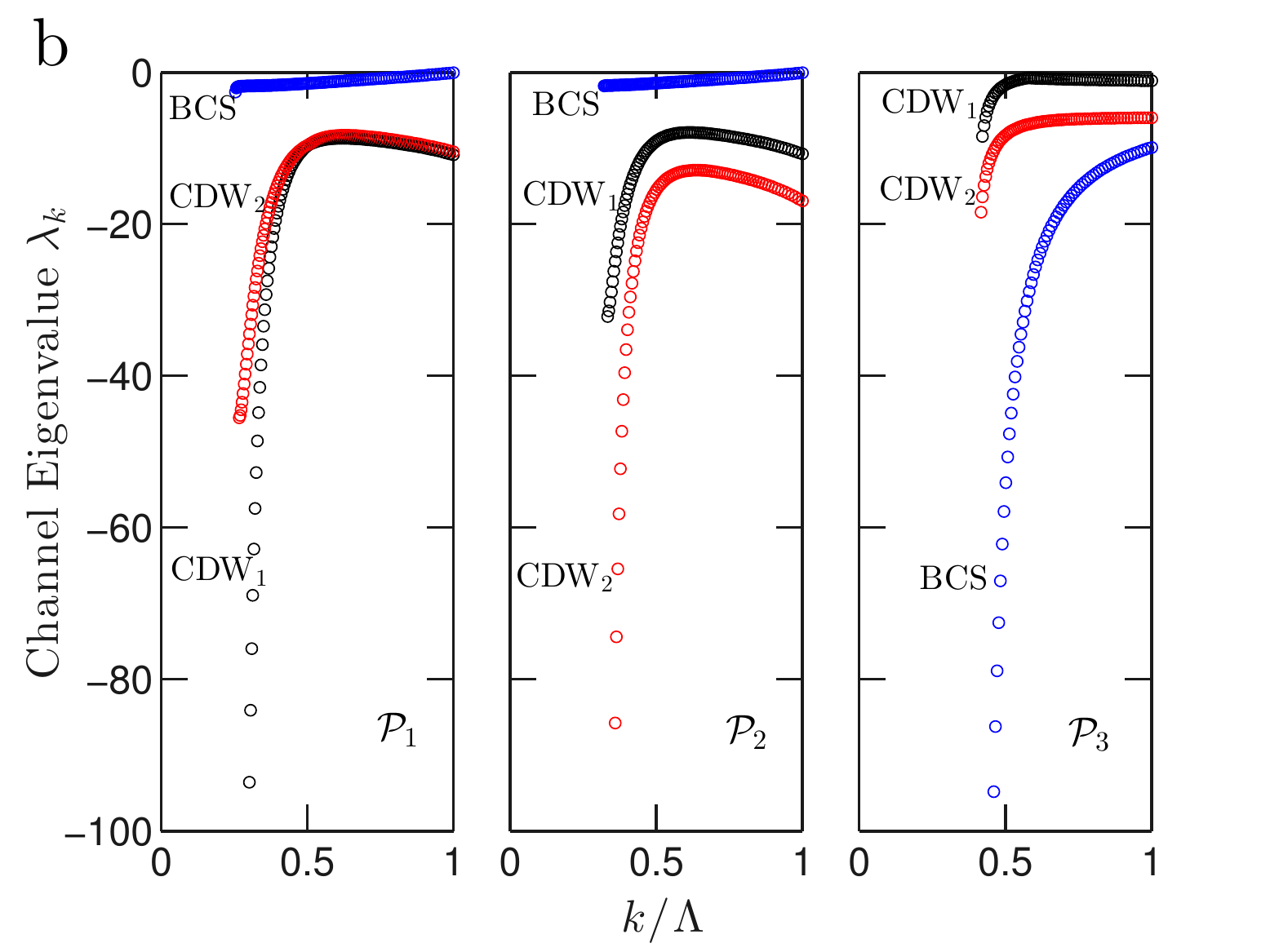}
  \includegraphics[width=0.47\textwidth]{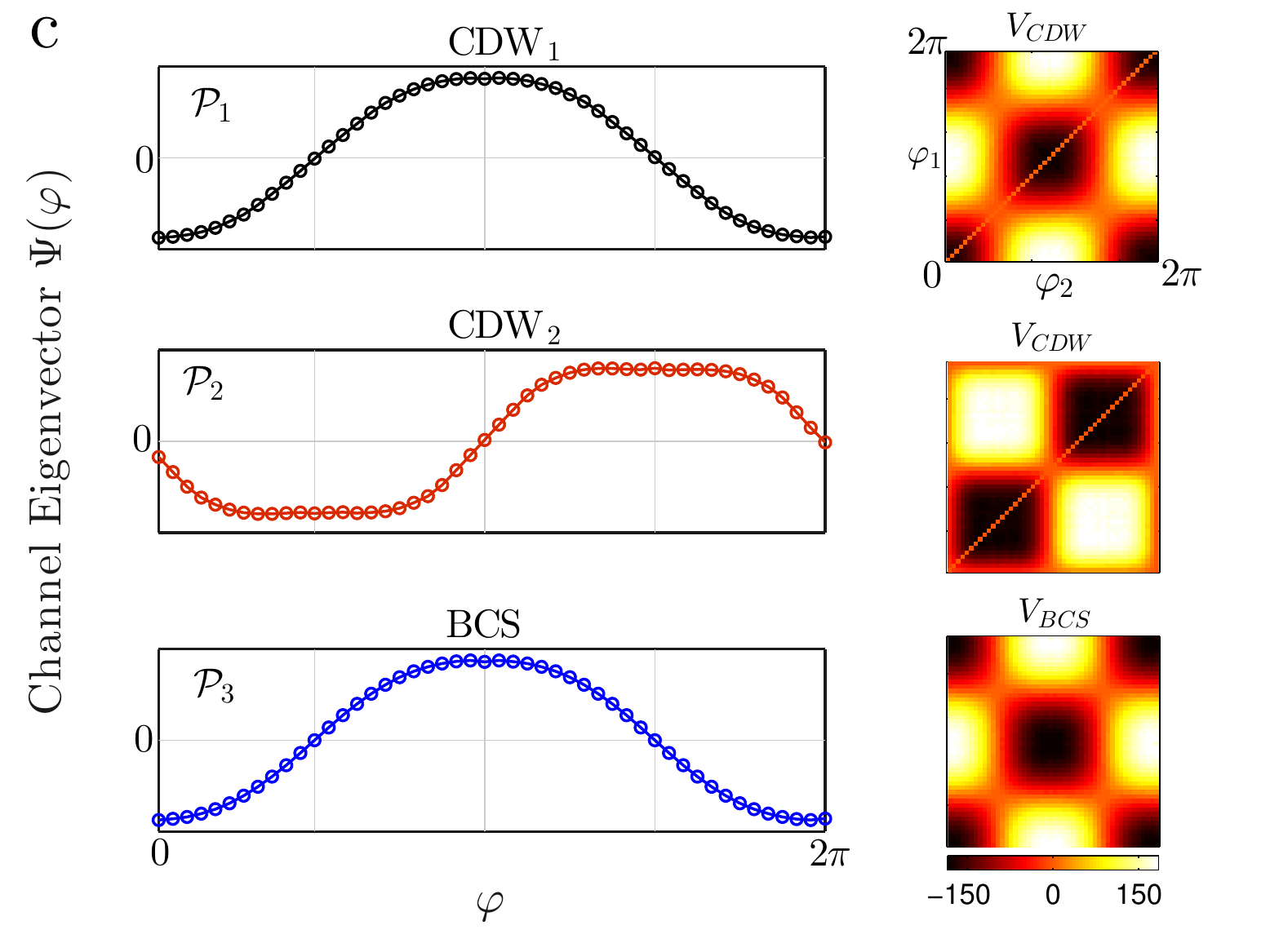}
  \caption{(Color online) a) The phase diagram of two-dimensional spinless
    dipolar Fermi gas at $T=0.01E_F$ predicted by FRG. It displays a Fermi
    liquid ($\mathrm{FL}$), a $p$-wave superfluid ($\mathrm{BCS}$) and two
    distinct density wave phases, $\mathrm{DW}_1$ and
    $\mathrm{DW_2}$.  The colormap shows the critical scale $\Lambda_c$ at
    which the vertex diverges (see the main text). Three representative points
    $\mathcal{P}_1$, $\mathcal{P}_2$ and $\mathcal{P}_3$ on the phase diagram
    are chosen to show the details of the FRG flow. b) The flows of the
    largest eigenvalue of $V_{BCS}$ and the two largest eigenvalues of
    $V_{DW}$ corresponding to the $\mathrm{DW}_1$ and $\mathrm{DW}_2$ order
    respectively. c) The eigenvectors of the leading instability and the
    corresponding channel matrices near the end of the flow.}
  \label{fig:PhaseDiagram}
\end{figure}

In this letter, we present a theory that meets the requirements (1)-(4) above.
It is based on functional renormalization group (FRG)
\cite{kopietz2010introduction,RevModPhys.84.299}, a powerful many-body
technique that gained considerable success in diverse systems including the
Hubbard model \cite{RevModPhys.84.299,zanchi}, the iron pnictides
\cite{RevModPhys.84.299,frg-adv-rev}, and ultracold quantum gases
\cite{PhysRevLett.97.030601,PhysRevB.75.174516,PhysRevA.89.053630,
  Tanizaki01042014,PhysRevLett.108.145301,PhysRevLett.110.155301}.  FRG can
accurately predict the leading instability of the interacting fermions without
a prior bias, making it the method of choice for problems with competing
orders. Our FRG analysis maps out the phase diagram of the 2D dipolar Fermi
gas at finite temperatures (Fig.~\ref{fig:PhaseDiagram}) which include a Fermi
liquid, a $p$-wave superfluid, and two distinct density wave phases outside
the ``collapse" region.  In particular, FRG reveals a DW phase with nontrivial
symmetry not found in the approaches summarized above.

Consider a continuum dipolar Fermi gas with chemical potential
$\mu=E_F=p_F^2/2m$, where $p_F$ is the Fermi momentum and $m$ is the mass of
the fermion. The dipole moment $ \mathbf{d}=d\hat{d}$ is aligned by external
electric or magnetic field in the direction
$\hat{d}=(\sin\theta,0,\cos\theta)$, i.e. titled toward the $x$ axis. The
interaction between two dipoles separated by a distance $\mathbf{r}$ is
$V_{dd}(\mathbf{r})=({d^2}/{r^3})[1-3(\hat{r}\cdot \hat{d})^2]$. Assume a
tight harmonic confinement of frequency $\omega_z$  in the $z$ direction, the
effective interaction for two fermions within the $xy$ plane has the form
\cite{PhysRevA.73.031602} 
\begin{equation}
  v(\mathbf{q})
  =  2\pi d^2 |\mathbf{q}|
  \left[(\hat{q}\cdot \hat{x})^2\sin^2\theta-\cos^2\theta\right]/\hbar,
  \label{eq:dipolarInteraction}
\end{equation}
where $\mathbf{q}=|\mathbf{q}|\hat{q}$ is the in-plane momentum transfer. Note
that Eq. \eqref{eq:dipolarInteraction} is only valid for $|\mathbf{q}|\leq
\Lambda<\hbar/l_0$ where $l_0=\sqrt{\hbar/m\omega_z}$ is the confinement
length. We introduce the dimensionless interaction strength ${g}=md^2p_F/\hbar^3
$ as the product of the typical interaction $2\pi d^2 p_F/\hbar$ and the
density of states $\nu = m/2\pi\hbar^2$. For brevity, we shall set $\hbar$ and
$k_B$ to be unity below.  Our goal is to find out which phase is stabilized
given the temperature $T$, the dipole tilting angle $\theta$, and the coupling
strength $g$. 

FRG implements Wilson's renormalization group for interacting
fermions (see for example Ref.~\onlinecite{rmp:shankar1994}) in an
exact and succinct fashion by flowing a generating functional, the average
effective action ${\Gamma}_k[\bar{\psi},\psi]$ where $\psi$ and $\bar{\psi}$
are the fermionic fields, as a sliding momentum scale $k$ is varied
\cite{RevModPhys.84.299,Wetterich199390,morris,Polchinski1984269}. Thermal and
quantum fluctuations on different scales are separated by a device called the
infrared regulator ${R}_k$ and dealt with successively at each scale. We adopt
Litim's optimized regulator \cite{litim2000optimisation}, 
\begin{equation}
  R_k(\xi_\mathbf{p})=
  [\mathrm{sign} (\xi_\mathbf{p}) k^2/2m- \xi_\mathbf{p}]
  \Theta(k^2/2m-|\xi_\mathbf{p}|),
\end{equation} 
where $\xi_\mathbf{p}=|\mathbf{p}|^2/2m-\mu$ is the bare dispersion,
$\mathbf{p}$ is the momentum within the $xy$ plane, and $\Theta$ is the
Heaviside step function. 
The evolution of ${\Gamma}_k$ obeys the exact flow equation
\cite{RevModPhys.84.299,Wetterich199390,Berges2002223}
\begin{equation}
\partial_k {\Gamma}_k[\bar{\psi},\psi]=-\frac{1}{2}\tilde\partial_k\mathrm{Tr}\ln 
\left[\hat{\Gamma}_k^{(2)}[\bar{\psi},\psi]+\hat{R}_k\right].
  \label{eq:WetterichEquation}
\end{equation}
Here $\hat{\Gamma}_k^{(2)}$ is the second order functional derivative of
${\Gamma}_k$ 
with respect to the fermionic fields, $\tilde\partial_k$ means $k$-derivative
only acting on $\hat{R}_k=i\hat{\sigma}_y R_k$, and Tr denotes integration
over the imaginary time $\tau\in[0,1/T]$, 
momentum $\mathbf{p}$, and
trace over $2\times 2$ matrices (denoted by hats) in the so-called superfield
space. The coarse-grained functional ${\Gamma}_k$ describes characteristic
correlations up to scale $k$, with all higher energy fluctuations integrated
out.  At the bare scale $\Lambda$, ${\Gamma}_{k=\Lambda}$ coincides with the
microscopic action. Thus, starting from the bare dispersion $\xi_\mathbf{p}$
and bare interaction $v(\mathbf{q})$ above and solving Eq.
\eqref{eq:WetterichEquation}, one can obtain an effective theory
${\Gamma}_{k\rightarrow 0}$ for the low-energy collective behaviors of the
interacting Fermi gas.

We expand ${\Gamma}_{k}$ up to quartic order of $\psi$ and $\bar{\psi}$,
\begin{equation}
\Gamma_k=\bar\psi_1 G^{-1}_k(p_1)\psi_1 
+\frac{1}{4}
  \Gamma^{(4)}_k(p_1,p_2,p_3)\bar\psi_4\bar\psi_3\psi_2\psi_1+...,
  \label{eq:VertexExpansion}
\end{equation}
with the short hand notation $\psi_i=\psi(p_i)$, $p=(p^0,\mathbf{p})$ where
the Matsubara frequency $p^0=(2n+1)\pi T$. By momentum conservation
$p_4=p_1+p_2-p_3$ in the four-point vertex $\Gamma^{(4)}_k$. Repeated indices
in Eq. \eqref{eq:VertexExpansion} are summed over, i.e. $T
\sum_{p^0}(2\pi)^{-2}\int {d^2\mathbf{p}}$ is implied for each $p_i$. The
inverse Green function is given by  $G^{-1}_k(p) =
{ip^0-\xi_\mathbf{p}^k-\Sigma_k(p)}$ with
$\xi_\mathbf{p}^k=\xi_\mathbf{p}+R_k(\xi_\mathbf{p})$ and $\Sigma_k$ is the
self-energy. Substituting Eq.~(\ref{eq:VertexExpansion}) into
Eq.\eqref{eq:WetterichEquation} and neglecting higher order vertices, we
obtain the coupled flow equations for $\Sigma_k(p)$ and
$\Gamma^{(4)}_k(p_1,p_2,p_3)$ \cite{Tanizaki01042014}.

Next we make a few standard approximations so that the flow equations become
numerically tractable. First, we neglect the $p^0$ dependence of
$\Gamma^{(4)}_k$ and set the  external Mastubara frequencies to be zero.
Second, we assume $\Gamma^{(4)}_k$ only depends on the direction (not the
magnitudes) of the momenta $\mathbf{\hat p}_1$, $\mathbf{\hat p}_2$ and
$\mathbf{\hat p}_3$. This approximation is similar to projecting the momenta
onto the Fermi surface in the widely used N-patch implementation of FRG for
fermions on lattices. Finally, we ignore the self-energy $\Sigma_k$.  With these
assumptions,  $\Gamma^{(4)}_k$ reduces to $\Gamma_k(\mathbf{\hat
  p}_1,\mathbf{\hat p}_2,\mathbf{\hat p}_3)$ which obeys the following flow
equation
\begin{align}
  &\partial_k\Gamma_k(\mathbf{\hat p}_1,\mathbf{\hat p}_2,\mathbf{\hat p}_3) 
  = 
  \nonumber\\
 &  \frac{1}{2}
 \int\frac{d\varphi_\mathbf{p}}{2\pi}
  \Gamma_k(\mathbf{\hat p}_1,\mathbf{\hat p}_2,\mathbf{\hat p}')
  \Gamma_k(\mathbf{\hat p}',\mathbf{\hat p},\mathbf{\hat p}_3)
  \tilde\partial_k
  \Pi^{+}_k(\mathbf{\hat p},\mathbf{\hat p}')
  \nonumber\\
   &+
  \int\frac{d\varphi_\mathbf{p}}{2\pi}
  \Gamma_k(\mathbf{\hat p}_1,\mathbf{\hat p},\mathbf{\hat p}')
  \Gamma_k(\mathbf{\hat p}',\mathbf{\hat p}_2,\mathbf{\hat p}_3)
  \tilde\partial_k
  \Pi^{-}_k(\mathbf{\hat p},\mathbf{\hat p}')
  \nonumber\\
   &-
  \int\frac{d\varphi_\mathbf{p}}{2\pi}
  \Gamma_k(\mathbf{\hat p}_1,\mathbf{\hat p}, \mathbf{\hat p}')
  \Gamma_k(\mathbf{\hat p}', \mathbf{\hat p}_2,\mathbf{\hat p}_4)
  \tilde\partial_k
  \Pi^{-}_k(\mathbf{\hat p},\mathbf{\hat p}'),
  \label{eq:flowGammap1p2p3}
\end{align}
where the angular integral is over $\varphi_\mathbf{p}$, the polar angle of the 2D
momentum $\mathbf{p}$. The three terms in Eq. \eqref{eq:flowGammap1p2p3} can
be represented diagrammatically as
\begin{center}
  \begin{tikzpicture}[scale=0.2,baseline=(current bounding box.center)]

    \def \x {-15}
    \def \y {2}
    \def \w {.7}
    \def \dver {6}
    \node [] at (\x+3.5,\y-5) {\scriptsize BCS};
    \draw[thick] (\x,\y) rectangle (\x+\w,\y+\w);
    \draw[thick] (\x+\w+\dver,\y) rectangle (\x+\dver+2*\w,\y+\w);
    \draw[middlearrow={stealth}] (\x-1,\y+\w+1) -- (\x,\y+\w);
    \node [left, black] at (\x-1,\y+\w+1.5) 
    {\scriptsize{1}};
    \draw[middlearrow={stealth}] (\x-1,\y-1) -- (\x,\y);
    \node [left, black] at (\x-1,\y-1.5) 
    {\scriptsize{2}};
    \draw[middlearrow={stealth}] 
    (\x+2*\w+\dver,\y+\w) -- (\x+2*\w+\dver+1,\y+\w+1);
    \node [right, black] at (\x+2*\w+\dver+1,\y+\w+1.5) 
    {\scriptsize{3}};
    \draw[middlearrow={stealth}] (\x+2*\w+\dver,\y) -- (\x+2*\w+\dver+1,\y-1);
    \node [right, black] at (\x+2*\w+\dver+1,\y-1.5) 
    {\scriptsize{4}};
    \draw[middlearrow={stealth reversed}] 
    (\x+\w+\dver,\y+\w) 
    to [out=145,in=35]
    (\x+\w,\y+\w);
    \node [above, black] at (\x+\w+.5*\dver,\y+\w+1.5)
    {$\mathbf{p}'$};
    \draw[middlearrow={stealth reversed}] 
    (\x+\w+\dver,\y) 
    to [out=-145,in=-35] 
    (\x+\w,\y);
    \node [below, black] at (\x+\w+.5*\dver,\y-1.5) 
    {$\mathbf{p}$};

    \node [] at (\x+11,\y+\w-0.5) {$+$};

    \def \x {0}
    \def \dver {6}
    \node [] at (\x+3.5,\y-5) {\scriptsize ZS};
    \draw[thick] (\x,\y) rectangle (\x+\w,\y+\w);
    \draw[thick] (\x+\w+\dver,\y) rectangle (\x+\dver+2*\w,\y+\w);
    \draw[middlearrow={stealth}] (\x-1,\y+\w+1) -- (\x,\y+\w);
    \node [left, black] at (\x-1,\y+\w+1.5) 
    {\scriptsize{1}};
    \draw[middlearrow={stealth}] (\x+2*\w+\dver+1,\y-1) -- (\x+2*\w+\dver,\y);
    \node [right, black] at (\x+2*\w+\dver+1,\y-1.5) 
    {\scriptsize{2}};
    \draw[middlearrow={stealth}] (\x,\y) -- (\x-1,\y-1);
    \node [left, black] at (\x-1,\y-1.5) 
    {\scriptsize{4}};
    \draw[middlearrow={stealth}] 
    (\x+2*\w+\dver,\y+\w) -- (\x+2*\w+\dver+1,\y+\w+1);
    \node [right, black] at (\x+2*\w+\dver+1,\y+\w+1.5) 
    {\scriptsize{3}};
    \draw[middlearrow={stealth reversed}] 
    (\x+\w+\dver,\y+\w) 
    to [out=145,in=35]
    (\x+\w,\y+\w);
    \node [above, black] at (\x+\w+.5*\dver,\y+\w+1.5)
    {$\mathbf{p}'$};
    \draw[middlearrow={stealth}] 
    (\x+\w+\dver,\y) 
    to [out=-145,in=-35] 
    (\x+\w,\y);
    \node [below, black] at (\x+\w+.5*\dver,\y-1.5) 
    {$\mathbf{p}$};

    \node [] at (\x+11,\y+\w-0.5) {$-$};

    \def \x {15}
    \node [] at (\x+3.8,\y-5) {\scriptsize ZS$'$};
    \draw[thick] (\x,\y) rectangle (\x+\w,\y+\w);
    \draw[thick] (\x+\w+\dver,\y) rectangle (\x+\dver+2*\w,\y+\w);
    \draw[middlearrow={stealth}] (\x-1,\y+\w+1) -- (\x,\y+\w);
    \node [left, black] at (\x-1,\y+\w+1) 
    {\scriptsize{1}};
    \draw[middlearrow={stealth}] (\x+2*\w+\dver+1,\y-1) -- (\x+2*\w+\dver,\y);
    \node [right, black] at (\x+2*\w+\dver+1,\y-1) 
    {\scriptsize{2}};
    \draw[middlearrow={stealth}] (\x,\y) -- (\x-1,\y-1);
    \node [left, black] at (\x-1,\y-1) 
    {\scriptsize{3}};
    \draw[middlearrow={stealth}] (\x+2*\w+\dver,\y+\w) 
    -- (\x+2*\w+\dver+1,\y+\w+1);
    \node [right, black] at (\x+2*\w+\dver+1,\y+\w+1) 
    {\scriptsize{4}};
    \draw[middlearrow={stealth reversed}] (\x+\w+\dver,\y+\w) 
    to [out=145,in=35]
    (\x+\w,\y+\w);
    \node [above, black] at (\x+\w+.5*\dver,\y+\w+1.5)
    {$\mathbf{p}'$};
    \draw[middlearrow={stealth}] (\x+\w+\dver,\y) 
    to [out=-145,in=-35] (\x+\w,\y);
    \node [below, black] at (\x+\w+.5*\dver,\y-1.5) 
    {$\mathbf{p}$};
  \end{tikzpicture}
  \nonumber
\end{center}
which are known as the BCS, the zero sound (ZS, or direct), and ZS$'$ (or
exchange) diagrams respectively. The internal momentum $\mathbf{p'} $ is given
by momentum conservation, $\mathbf{p'}=\mathbf{p}_1+\mathbf{p}_2-\mathbf{p}$
for BCS channel, $\mathbf{p'}=\mathbf{p}+\mathbf{p}_2-\mathbf{p}_3$ for ZS
channel, and $\mathbf{p'}=\mathbf{p}+\mathbf{p}_1-\mathbf{p}_3$ for ZS$'$
channel.  The polarization bubbles in the particle-particle channel,
$\Pi^{+}_k$, and the particle-hole channel, $\Pi^{-}_k$, are given by 
\begin{equation}
  \Pi^{\pm}_k(\mathbf{\hat p},\mathbf{\hat p}')=
 T\sum_{p^0}
  \int\frac{d|\mathbf{p}|^2}{4\pi}
  \frac{1}{( ip^0-\xi_\mathbf{p}^k)}
  \frac{1}{(\mp ip^0-\xi_\mathbf{p'}^k)}.
  \label{eq:bubbles}
\end{equation}
After performing the Matsubara summations analytically and
the radial integral numerically in Eq.~\eqref{eq:bubbles}, 
we find that
the main contribution to $\tilde\partial_k \Pi^{\pm}_k$ comes from the
intersection area of two annuli in momentum space,  
\begin{equation}
  \tilde\partial_k
  \Pi^{\pm}_k
  = \pm \nu\int d\xi
  f_{k}(\xi,\pm\xi')
  \Theta(\tilde{k}^2-|\xi|)\Theta(\tilde{k}^2-|\xi'|),
  \label{eq:PartialPi}
\end{equation}
where $\tilde{k}=k/\sqrt{2m}$, 
$\xi= \xi_\mathbf{p}$, $\xi'=\xi_\mathbf{p'}$, and the weight function
$
f_{k}(\xi,\xi')
=[
  x_k/2\cosh^{2}(x_k/2)
  -\tanh (x_k/2) ]/{\tilde{k}^3}
$
for $\xi \xi'>0 $
and
$ f_{k}(\xi,\xi')
= -x_k^2
 \tanh (x_k)/[2\cosh^{2}(x_k)  {\tilde{k}^3}]
$
for 
$
 \xi\xi'<0
$ in terms of $x_k=k^2/2mT$. The vertex function at scale $\Lambda$ is 
the antisymmetrized bare interaction,
\begin{equation}
  \Gamma_\Lambda(\mathbf{\hat p}_1,\mathbf{\hat p}_2,\mathbf{\hat p}_3) 
  =
  \frac{1}{2}
  \left[
    v(\mathbf{\hat p}_3-\mathbf{\hat p}_1) -
    v(\mathbf{\hat p}_2-\mathbf{\hat p}_3)
  \right].
  \label{eq:barevertex}
\end{equation}
Eqs. \eqref{eq:flowGammap1p2p3}-\eqref{eq:barevertex} are the main analytical
results of our paper.

We solve the flow Eq. \eqref{eq:flowGammap1p2p3} with the initial condition
\eqref{eq:barevertex} numerically by discretizing the sliding scale
$k\in [0,\Lambda]$ and the polar angle $\varphi_\mathbf{p}\in [0,2\pi]$. We
choose $\Lambda=0.4p_F$, which corresponds to an energy scale much larger than
$T$~\cite{PhysRevB.52.13487,PhysRevB.57.1444}, and an angular grid of $N=48$
patches. The evolution of $\Gamma_k(\mathbf{\hat p}_1,\mathbf{\hat
  p}_2,\mathbf{\hat p}_3)$, which contains $N^3=110592$ running coupling
constants, is monitored as $k$ is reduced from $\Lambda$ toward 0.  To
identify the many-body instabilities, we introduce the standard coupling
matrices in various channels \cite{PhysRevB.80.064517,frg-adv-rev}, e.g.
$V_Q(\mathbf{\hat p}_1,\mathbf{\hat p}_2)$ by setting
$\mathbf{p}_3=\mathbf{p}_1+\mathbf{Q}$ in $\Gamma_k$ for given $\mathbf{Q}$.
It turns out the leading instability occurs either in the particle-particle
channel with  the coupling matrix $V_{BCS}(\mathbf{\hat p}_1,\mathbf{\hat
  p}_2)$=$[\Gamma_k(\mathbf{\hat p}_1,-\mathbf{\hat p}_1,\mathbf{\hat
  p}_2)+\Gamma_k(\mathbf{\hat p}_2,-\mathbf{\hat p}_2,\mathbf{\hat p}_1) ]/2$
or in the particle-hole channel with the coupling matrix $V_{DW}(\mathbf{\hat
  p}_1,\mathbf{\hat p}_2)$=$[\Gamma_k(\mathbf{\hat p}_1,\mathbf{\hat
  p}_2,\mathbf{\hat p}_1)+\Gamma_k(\mathbf{\hat p}_2,\mathbf{\hat
  p}_1,\mathbf{\hat p}_2)]/2$. 

Each channel coupling matrix $V$ is diagonalized to find its eigenvalues and
eigenvectors, $   \int\frac{d\varphi_2}{2\pi} V(\varphi_1,\varphi_2)
\Psi(\varphi_2)=\lambda_k\Psi(\varphi_1)$, where we have used the unit vector
$\mathbf{\hat p}_i$ and its corresponding polar angle $\varphi_i$
interchangeably.  For given $(\theta,g,T)$, if $\Gamma_k$ develops no singular
behavior as $k$ is reduced, the gas is in the normal phase. If some
$\lambda_k$ values diverge, the Fermi liquid is unstable. The most diverging
$\lambda_k$ points to the channel in which the leading instability occurs,
while its eigenvector $\Psi(\varphi)$ yields the symmetry of the incipient
long-range order. We define the critical scale $\Lambda_c$ as the value of $k$
when the largest element of $\Gamma_k(\mathbf{\hat p}_1,\mathbf{\hat
  p}_2,\mathbf{\hat p}_3)$ exceeds $10^2E_F$. It serves as a rough estimation
of the critical temperature $T_c$ of the corresponding broken symmetry phase.

Identifying the instability in the particle-hole channel for a Fermi gas with
circular Fermi surface requires some care. $V_{DW}$ may look like the $\mathbf{Q}\rightarrow 0$ limit of
$V_Q$. It however can be equivalently viewed as two
fermions exchanging momentum, $(\mathbf{p}_1, \mathbf{p}_2)
\rightarrow (\mathbf{p}_2, \mathbf{p}_1)$, with a finite momentum transfer
$\mathbf{Q'}=\mathbf{p}_2-\mathbf{p}_1$, due to the antisymmetry of the vertex. In particular, scattering across
the Fermi surface, $(\mathbf{p}_1, -\mathbf{p}_1)
\rightarrow (-\mathbf{p}_1, \mathbf{p}_1)$, drives a DW (stripe) order
with ordering wave vector $|\mathbf{Q}'|=2p_F$ \cite{PhysRevB.90.155102}. Within
our FRG scheme, we find that while the DW order does
have a dominant $\mathbf{Q'}$ component, it cannot be characterized by 
a single wave vector~\cite{PhysRevA.84.063633}.

Fig.~\ref{fig:PhaseDiagram}a shows the phase diagram of 2D dipolar Fermi gas
for $T=0.01E_F$. It features four phases: the Fermi liquid (FL), the BCS
superfluid phase dominated by $p_x$-wave symmetry, and two distinctive DW
phases. For completeness, we also indicated the ``collapse" region obtained by the same procedure as 
Ref. \cite{PhysRevA.84.063633}.  Three
points on the ($\theta, g)$ plane, $\mathcal{P}_1$, $\mathcal{P}_2$ and
$\mathcal{P}_3$, are chosen to represent the DW$_1$, DW$_2$, and BCS phase
respectively. Their corresponding $\lambda_k$, $\Psi(\varphi)$, and
$V(\varphi_1,\varphi_2)$ are shown in
Fig.~\ref{fig:PhaseDiagram}b-\ref{fig:PhaseDiagram}c.
 
Several features of Fig.~\ref{fig:PhaseDiagram}a are in qualitative agreement
with previous phase diagrams in Refs.
\cite{PhysRevLett.101.245301,PhysRevA.84.063633,PhysRevLett.108.145304,PhysRevB.91.224504}.
For example, the BCS phase emerges beyond a critical dipole tilting angle
$\theta_c\approx0.26 \pi$, while the DW order only develops beyond a critical
coupling, e.g. $g_c=1.1$ for $\theta=0$.  
%
Fig.~\ref{fig:PhaseDiagram}a shows that the BCS phase undergoes a direct transition to DW$_2$ instead of
through an intermediate, coexisting phase. Within our implementation of FRG, 
the leading instability always occurs either in the BCS or the DW channel. The case of degenerately diverging $\lambda_k$ in
both channels is never observed, so there is no evidence for a supersolid phase. We can identify DW$_2$ as the stripe
density wave discussed previously with a density modulation along the $y$ axis
with period $\sim\hbar/2p_F$.  As shown in Fig.~\ref{fig:PhaseDiagram}c, the
effective interaction $V_{DW}$ is repulsive and diverging near the region
$\mathbf{\hat{p}}_1\sim -\mathbf{\hat{p}}_2\sim\hat{y}$, favoring the dominant
ordering wave vector $\mathbf{Q}'=2p_F \hat{y}$. Note that $\Psi(\varphi)$
vanishes for $\varphi=0$ and $\pi$, i.e. there is no modulation along the $x$
direction.  
\begin{figure} 
  \centering 
  \includegraphics[width=0.45\textwidth]{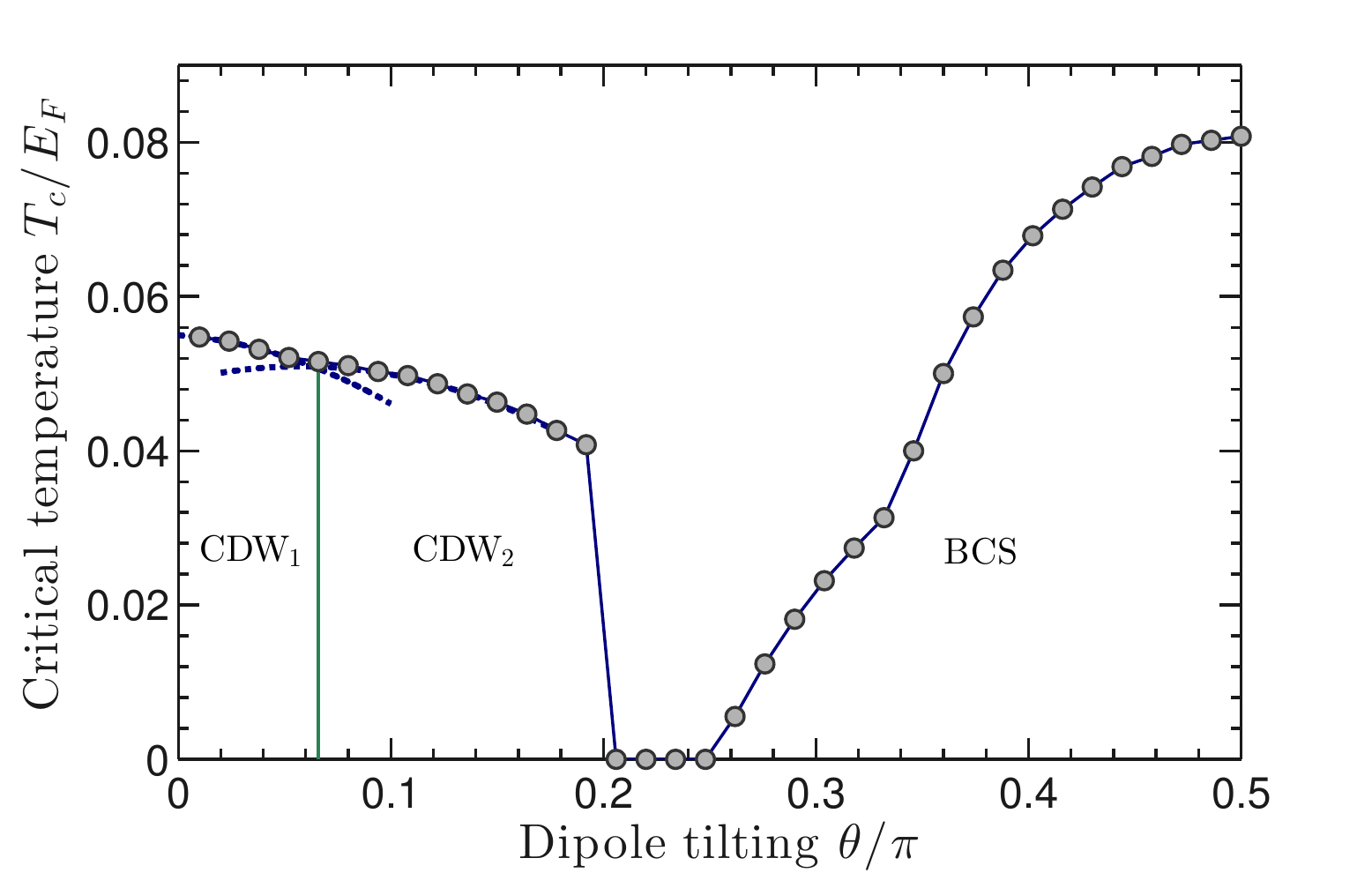}
  \caption{The critical temperature $T_c$ as a function of the dipole tilting
    angle $\theta$ for fixed interaction $g=1.5$ estimated from the FRG flow. Dashed
    lines are the extensions of 
    the fits to the data points in the
    $\mathrm{DW}_1$ and $\mathrm{DW}_2$ region, resepectively.} 
  \label{fig:tc} 
\end{figure}

Our most significant finding is the DW$_1$ phase seen for smaller dipolar
tilting angles $\theta$.  From the eigenvector $\Psi(\varphi)$, it is clear
that the two DW phases are roughly related to each other by a $\pi/2$
rotation. The density modulation is thus along the $x$ axis in the DW$_1$
phase. In contrast to DW$_2$, the DW$_1$ order is not expected from the bare
interaction $v(\mathbf{q})$ which is almost isotropic for small $\theta$. Only
under the FRG flow does the effective interaction vertex $V_{DW}$ become
increasingly anisotropic and drastically different from the bare interaction.
As $k$ is reduced, the renormalized interaction for two fermions with
$\mathbf{\hat{p}}_1\sim -\mathbf{\hat{p}}_2\sim\hat{x}$ grows dominantly
repulsive and eventually diverges. While the translational symmetry breaking
along the $x$ direction is counter-intuitive, a DW order along $\hat{y}$ in
the limit of $\theta\rightarrow 0$, as previously believed, seems implausible
because all directions of $\mathbf{Q}'$ are energetically degenerate. 
%

Single channel renormalization group \cite{rmp:shankar1994} and RPA are widely
used in the study 2D Fermi liquids, and their deficiencies have been noticed
\cite{PhysRevB.57.1444,dupuis}. Neglecting the ZS$'$ channel (as in RPA) will
violate the antisymmetry in the forward scattering vertex
\cite{PhysRevB.57.1444}. Strong interference between ZS and ZS$'$ near the
zero angle $\mathbf{\hat{p}}_1\sim \mathbf{\hat{p}}_2$ leads to small angle
anomalies and invalidates the ladder approximation \cite{PhysRevB.57.1444}.
The interference between the BCS and ZS channel (relevant to the putative
supersolid phase) become important when $\mathbf{\hat{p}}_1\sim
-\mathbf{\hat{p}}_2$ \cite{dupuis,rmp:shankar1994}. The FRG approach described here is
capable of describing these subtle interplays between the BCS, ZS, and ZS$'$
channels, and the antisymmetry of $\Gamma_k$ is respected throughout.

Now we comment on the requirements to observe the broken symmetry phases
experimentally. Fig.~\ref{fig:PhaseDiagram}a suggests that to access the BCS
phase at $T\sim 0.01E_F$, which is much lower than the temperatures $T\sim 0.2 E_F$ achieved
for Dy \cite{PhysRevLett.108.215301} and Er \cite{PhysRevLett.112.010404} gases, one needs go to relatively strong interactions $g>0.5$. For the 
NaK gas reported in Ref. \cite{PhysRevLett.114.205302}, $g$ is on the order of 1.34
with $d=0.8$Debye and area density of $4\times 10^{7}$cm$^{-2}$. In principle,
$d$ can be further increased to 2.7Debye, giving a ten-fold increase in $g$.
At higher temperatures, the phase boundaries are shifted to the right by
thermal fluctuations, so stronger interactions are required to reach the BCS
and DW phases.  The (color-coded) $\Lambda_c$ values in
Fig.~\ref{fig:PhaseDiagram}a provides a rough guide for the $T_c$.  Note the
$T_c$ of the BCS phase is typically higher than that of the DW phases.  A
more accurate estimation of $T_c$ is shown in Fig.~\ref{fig:tc} for fixed $g=1.5$ as a
function of $\theta$.  It is obtained by solving the flow equation at
different $T$ for given $(\theta,g)$ and identifying $T_c$ as the temperature
at which a divergence in $\Gamma_k$ develops.  We observe that $T_c$ can be 5
to 10 percent of $E_F$.

Our instability analysis is confined within the normal phase,
it does not directly describe the broken symmetry phases including the proliferation of topological defects
which tends to suppress $T_c$ to values much lower than estimated here. 
The Kosterlitz-Thouless transition of the stripe (DW$_2$) phase has been described in Ref.  \cite{wu2015liquid}. 
It would be interesting to perform a similar analysis of all candidate phases found here. 
The accuracy of our results can be improved by
including the full momentum dependence in $\Gamma^{(4)}_k$ and $\Sigma_k$ in
the FRG flow.  For example, we expect the Fermi surface
deformations \cite{Aikawa19092014,miyakawa,wu-sarma,yi-FS} due to the
self-energy correction may slightly reduce the region of the $\mathrm{DW}_1$ phase and 
enhance the stability of the supersolid.
Such calculation is numerically much more
demanding and reserved for future work. We hope the results reported here can
stimulate further application of FRG to interacting Fermi gases 
including spin 1/2 dipolar Fermi gases \cite{
yi-wu-review,PhysRevA.87.043604,PhysRevB.92.081106}.

\begin{acknowledgments}
E.Z. is grateful to Xiaopeng Li, Ludwig Mathey, Shan-Wen Tsai, and Ryan Wilson
for many helpful discussions. This work is supported by the NSF Grant No.
PHY-1205504 and the AFOSR Grant No. FA9550-16-1-0006. A.K. is also supported
by U.S. ARO Grant No. W911NF-11-1-0230.
\end{acknowledgments}
\bibliography{refs,coldatoms}
\end{document}